\documentclass[12pt]{article}
\pdfoutput=1
\usepackage{amsmath,amssymb,amsfonts,epsfig,graphicx,euscript}%
\usepackage[pdftex,bookmarks,bookmarksnumbered,linktocpage,pdfstartview=FitH]{hyperref}
\hypersetup{colorlinks,%
citecolor=red,%
filecolor=blue,%
linkcolor=blue,%
urlcolor=blue,%
pdftex}
\usepackage[all]{hypcap}
\numberwithin{equation}{section}
\usepackage{cite}

\newcommand{\bse}{\begin{subequations}}
\newcommand{\ese}{\end{subequations}}
\newcommand{\be}{\begin{equation}}
\newcommand{\ee}{\end{equation}}
\newcommand{\bea}{\begin{eqnarray}}
\newcommand{\eea}{\end{eqnarray}}
\newcommand{\ba}{\begin{array}}
\newcommand{\ea}{\end{array}}

\begin{document}
\hfill%
\vbox{
    \halign{#\hfil        \cr
           IPM/P-2013/019\cr
                     }
      }
\vspace{1cm}
\begin{center}
{ \Large{\textbf{Probe Branes Thermalization in External Electric and Magnetic Fields}}}
\vspace*{1cm}
\begin{center}
{\bf M. Ali-Akbari$^{a,}$\footnote{aliakbari@theory.ipm.ac.ir}, H. Ebrahim$^{b,}$\footnote{hebrahim@ipm.ir}, Z. Rezaei$^{a,c,}$\footnote{z.rezaei@aut.ac.ir}}\\%
\vspace*{0.4cm}
{\it {$\ ^{a}$School of Particles and Accelerators,$\ ^{b}$School of Physics,\\ 
Institute for Research in Fundamental Sciences (IPM),\\
P.O.Box 19395-5531, Tehran, Iran}}  \\
{\it {$\ ^{c}$Department of Physics, University of Tafresh,\\
P.O.Box 39518-79611, Tafresh, Iran}}  \\

\vspace*{0cm}
\end{center}
\end{center}

\vspace{0cm}
\begin{center}
\textbf{Abstract}
\end{center}
We study thermalization on rotating probe branes in $AdS_5\times S^5$ background in the presence of constant external electric and magnetic fields. In the AdS/CFT framework this corresponds to thermalization in the flavour sector in field theory. The horizon appears on the worldvolume of the probe brane due to its rotation in one of the sphere directions. For both electric and magnetic fields the behaviour of the temperature is independent of the probe brane dimension. We also study the open string metric and the fluctuations of the probe brane in such a set-up. We show that the temperatures obtained from open string metric and observed by the fluctuations are larger than the one calculated from the induced metric.

\newpage

\tableofcontents

\section{Introduction and result}
Quark-gluon plasma (QGP), as a new phase of quantum chormodynamics,
is produced in relativistic heavy ion collisions at RHIC and,
these days, at LHC. Immediately after the collision, the QGP
is out of equilibrium and after a very short time, usually called ``thermalization time'', the 
system gets thermalized \cite{Heinz:2004pj,Luzum:2008cw}. This phenomena is known as ``rapid thermalization'' in the literature. 
In fact at this stage the QGP can be described by viscous hydrodynamics equations. Using the
hydrodynamics equations describing a fluid with a small value of
shear viscosity over entropy density, $\eta/s$, one can reproduce
experimental data \cite{Shuryak:2003xe,Shuryak:2004cy}. This smallness indicates 
the strongly coupled nature of the plasma produced at the collision. Therefore it seems that perturbative methods are not very much reliable to obtain physical
quantities related to various properties of the QGP. Apart from strong coupling, out of
equilibrium characteristic of the system makes it more complicated to study
the process of thermalization during this short period of time.

The AdS/CFT correspondence can be considered as a framework to study out of equilibrium
strongly coupled systems. This
correspondence states that type IIB string theory on $AdS_5\times
S^5$ background, which describes near horizon geometry of a stack of
$N_c$ extremal D3-branes, is dual to the four-dimensional
super-conformal Yang-Milles theory with the gauge group
$SU(N_c)$ \cite{Maldacena}. The correspondence is more applicable in the
limit of large $N_c$ and large 't Hooft coupling,
$\lambda=g_{YM}^2N_c$ where $g_{YM}$ is the gauge theory coupling
constant. In these limits it is well-known that a strongly coupled
SYM is dual to the IIb supergravity which is the low energy
effective theory of IIB superstring theory. This correspondence has
also been generalized to the thermal SYM theories. As a result a
strongly coupled thermal SYM theory corresponds to the supergravity
in an AdS-Schwarzschild black hole background where SYM theory
temperature is identified with the Hawking temperature of the AdS
black hole \cite{Witten:1998zw}.

Matter fields (or quarks) in the fundamental representation of the
$SU(N_c)$ gauge group can be added to the original SYM theory by
introducing $N_f$ flavour D$p$-branes in the probe limit on the
gravity side \cite{Karch:2002sh}. The probe limit means that the number of
flavour D$p$-branes are much smaller than the number of D3-branes or
in other words the D$p$-branes do not modify the geometry. This will add ${\cal{N}}=2$ fundamental hypermultiplets, which live on the defect, to the 4-dimensional ${\cal{N}}=4$ conformal SYM theory. 

Thermalization in field theory happens after an injection of
energy into the system. The energy injection can be done in two ways: by explicitly adding a time-dependent source to the field theory Lagrangian or by introducing a time-dependent coupling in the field theory. According to the AdS/CFT dictionary the first one corresponds to horizon formation in the bulk and so thermalization in the gluon sector \cite{Chesler:2008hg,Bhattacharyya:2009uu}. The second one results in horizon formation on the probe brane corresponding to thermalization in the flavour sector \cite{Das:2010yw,Das:2011nk,Hashimoto:2010wv,AliAkbari}. In the second set-up the background temperature
is always kept fixed, indicating that we are working in the probe
limit. Therefore in the field theory picture the flavour sector is thermalizaed to a temperature different from the bulk or the gluon sector. Considering the backreaction of the flavour branes will lead to horizon formation in the bulk and therefore thermalizing the gluon sector.  


To understand thermalization in the flavour sector an interesting set-up has been introduced in \cite{Das:2010yw} in which a probe rotating D$p$-brane has been studied in the $AdS_{d+2}\times S^q$ backgrounds. In fact the probe D$p$-brane is rotating, by an external force, with a constant angular velocity in the largest
circle in $S^q$. The solution to the equations of motion of the worldvolume theory for different D$p$-branes results in
a non-trivial time-dependent classical solution. This represents
a time-dependent mass or coupling in the defect field theory which effectively produces a thermal state. The rotation causes the induced metrics
on the probe branes to have a horizon and therefore the degrees of freedom living on the probe brane worldvolume will see a nonzero Hawking temperature. This temperature is equal to $\frac{\sqrt{p}\omega}{2 \pi}$ for $p\le d+1$ where $\omega$ is angular velocity. Therefore the flavour sector is thermalized to a temperature different from the SYM theory and the system is basically out of equilibrium.

Based on the set-up explained in the previous paragraph \cite{Das:2010yw}, in this paper we consider a general profile for the spinning D$p$-branes and would like to study the effect of external electric and magnetic fields on the temperature of the thermalized flavour sector and the profile of the probe branes. For non-rotating D$p$-branes such studies have been extensively done in the literature. For more information look at \cite{Erdmenger:2007bn} and their references. Our observations can be summarized as follows. One can find more details throughout the paper. 

In the case where the magnetic field is nonzero the flavour sector always gets thermalized due to the rotation but the temperature decreases as the magnetic field is raised until it reaches a constant value. This constant value depends linearly on the angular velocity, $\omega$, for $AdS_5\times S^5$ background in Poincare' coordinate. We also observe that for fixed magnetic field the temperature increases with raising the angular velocity until it behaves linearly with respect to it, independent of the value of the magnetic field.

For the case where the electric field is nonzero on the probe brane the above results are modified. In contrast to the magnetic field case the flavour sector does not thermalize until the angular velocity reaches a minimum value given by the electric field. After that the temperature increases by raising angular velocity and for large $\omega$, similar to the magnetic field case, it behaves linearly with respect to it. We also show that the flavour sector temperature decreases as one increases the electric field until it reaches zero. Therefore in contrast to the magnetic field case there exists a maximum value for the electric field fixed by $\omega$ after which the flavour sector will not thermalize.

We have also studied the temperature obtained from the open string metric \cite{Seiberg:1999vs} for nonzero electric field. It is different and larger than the temperature obtained from the induced metric. We have intuitively explained in section \ref{fluctuationsection} the reason behind this observation. The scalar fluctuations of the probe brane along the transverse directions have been also discussed. The metric in the action for fluctuations is different from the open string metric by a scaling factor and thus results in a different temperature. Therefore we will observe that since the system is out of equilibrium, due to the rotation and nonzero electric field in the defect theory, different  constituents of the flavour sector see different temperatures which become the same in the $E\rightarrow 0$ limit.

\section{Worldvolume horizon on the rotating D$p$-branes }
In this section we describe the set-up of the problem in more details. We are interested in studying
thermalization on probe D$p$-branes. As explained in the introduction
this can be done by rotating the probe brane in one of the
transverse directions to the probe brane, along one of the sphere
directions. The rotation will cause the induced metric on the probe
brane to have a horizon and therefore the fundamental matter sector
will feel a nonzero temperature although the background temperature
is zero. 

The background metric in which the probe brane is rotating is an asymptotically
$AdS_{d+2}\times S^q$ metric which has the general form
\be\label{background} %
 ds^2=g_{tt}dt^2+g_{rr}dr^2+\sum_{i=1}^d
 g_{ii}dx_i^2+g_{\theta\theta}d\theta^2+g_{\phi\phi}d\phi^2+g_{ss}d\Omega^2_{q-2}.
\ee %
The low energy effective action for a D$p$-brane embedded in the above background is described by Dirac-Born-Infeld (DBI) and Chern-Simons (CS) actions %
\be\label{action}\begin{split}  %
 S&=S_{DBI}+S_{CS},\cr
 S_{DBI}&=-\mu_{p}\int d^{p+1}\xi\; e^{-\phi}\sqrt{-\det (G_{ab}+B_{ab}+2\pi \alpha'
 F_{ab})},\cr
 S_{CS}&=\mu_{p}\int P[\Sigma C^{(n)}e^{B}]e^{2\pi \alpha' F},
\end{split}\ee %
where $\xi$'s are worldvolume coordinates and $\mu_{p}^{-1}=(2\pi)^pl_s^{p+1}g_s$ is the D$p$-brane tension.  $G_{ab}$ and  $B_{ab}$ represent the induced metric and Kalb-Ramond field on the D$p$-brane defined as
\be\label{induced metric}%
G_{ab}=g_{MN}\partial_a X^M \partial_b X^N,
\ee %
\be\label{Kalb-Ramond}%
B_{ab}=B_{MN}\partial_a X^M \partial_b X^N,
\ee %
where the capital indices represent spacetime coordinates. $g_{MN}$ is the background metric \eqref{background} and $B_{MN}$ is zero for all cases we consider in this paper. $F_{ab}$ is the field strength of the gauge field living on the D$p$-brane. In the CS action, $C^{(n)}$ is a $n$-form field and $P[...]$ is the pullback of bulk spacetime tensors on the D$p$-brane world-volume. The contribution of the CS action is also zero for the cases we discuss in the paper.

We start with $p=2$ and $3$ and discuss the D7-brane in the next section. In order to embed the D$p$-brane we use the static gauge where the brane expands along $t$, $r$, $x_{1}$, ... and $x_{p-1}$ directions. We are interested in classical configurations where the probe brane is rotating in one of the sphere coordinates, which we assume to be $\phi$. The ansatz we choose for the transverse directions has the form
\be\label{configuration} %
 \phi(t,r)=\omega t+g(r),\ \ \theta(r),
\ee %
and the rest of them are set to zero. It is well known that adding the flavour D-branes in the probe limit to the bulk is equivalent to introducing matter fields in the fundamental representation of the corresponding gauge group on the gauge
theory side \cite{Karch:2002sh}.
As it has been discussed in \cite{Das:2010yw}, due to rotation these matter
fields have the mass proportional to $e^{-iwt}$. Moreover as we will show
the Hawking temperature defined on the D-branes is sensitive to the classical solution for
$\theta(r)$. 

The induced metric on the probe brane is then given by
\be %
 ds_{ind}^2=G_{tt}dt^2+2G_{tr}dtdr+G_{rr}dr^2+\sum_{i=1}^{p-1}
 G_{ii}dx_i^2,
\ee %
where
\be\begin{split} %
 G_{tt}&=g_{tt}+g_{\phi\phi}\omega^2, \cr %
 G_{tr}&=g_{\phi\phi}\omega g'(r), \cr %
 G_{rr}&=g_{rr}+g_{\phi\phi}g'(r)^2+g_{\theta\theta}\theta'(r)^2, \cr%
 G_{ii}&=g_{ii}, \cr %
\end{split}\ee %
and all other components are zero. 

In order to obtain the temperature on the probe brane one can use the following change of coordinate %
\be %
 d\tau=dt+k(r)g'(r)dr,
\ee %
where %
\be %
 k(r)=\frac{g_{\phi\phi}\omega}{g_{tt}+g_{\phi\phi}\omega^2}.
\ee %
It is then straightforward to write the induced metric as %
\be %
 ds_{ind}^2=f(r)d\tau^2+h(r)dr^2+\sum_{i=1}^{p-1}g_{ii}dx_{i}^2,
\ee %
where %
\bea\label{horizon2} %
 f(r)&=&g_{tt}+g_{\phi\phi}\omega^2, \cr
 h(r)&=&\frac{(g_{rr}+g_{\theta\theta}\theta'(r))(g_{tt}+g_{\phi\phi}\omega^2)+g_{tt}g_{\phi\phi}g'(r)^2}{g_{tt}+g_{\phi\phi}\omega^2}.
\eea %
Therefore the horizon on the D$p$-brane is located at $r=r_h$ which is given by %
\be\label{horizon}
f(r_h)=h^{-1}(r_h)=0,
\ee%
after replacing $\theta(r)$ and $g(r)$ by their classical solutions.
So the temperature on the probe brane is
\be\label{temperature} %
 T=\frac{1}{4\pi}\sqrt{- \frac{f'(r_h)}{h'(r_h)}}.
\ee %
Let's assume the bulk metric does not have a horizon and the bulk temperature is zero.
Therefore although the bulk temperature is zero there will be a nonzero
temperature on the probe brane due to its rotation \cite{Das:2010yw}. This shows that the flavour sector
or the degrees of freedom on the probe brane feel the non-zero temperature proportional
to the angular velocity, $\omega$, while the gluon sector is at zero temperature.

In the following subsections we will add nonzero external electric and magnetic fields
to different probe branes and will study how it might affect the temperature observed
on them. This can not be done in a general closed form therefore we will
do it case by case for D2 and D3-branes.

\subsection{D2-brane}\label{D2brane}
The simplest set-up in which one can study the effect of the external electric field is the D2-brane
embedded in the background \eqref{background}. The
D2-brane expands along $(t, r, x_{1})$ directions of $AdS$ space and
rotates in the $\phi$ direction as it was assumed in \eqref{configuration}. We consider a constant external electric field on the brane where its field
strength is in the $t$ and $x_{1}\equiv x$ directions, \textit{i.e.} $F_{tx}=E$.
Using \eqref{action} the Lagrangian density will be
\be\label{lagrangian}\begin{split}
 {\cal{L}}=&\bigg{[}-\left(g_{tt}g_{xx}+(2\pi\alpha')^2 E^2\right)\left(g_{rr}
 +g_{\phi\phi}g'(r)^2+g_{\theta\theta}\theta'(r)^2\right)\cr
 &-g_{xx}g_{\phi\phi}\omega^{2}(g_{rr}+g_{\theta\theta}\theta'(r)^2)\bigg{]}^{1/2}.
\end{split}\ee %
Notice that apart from the first parenthesis under the square root in
the above equation, the second parenthesis and the second term in the
Lagrangian are always positive. As a result the reality
condition of the action requires that %
\be\label{reality} %
|g_{tt}|g_{xx}>(2\pi\alpha')^2 E^2.
\ee %
For the values of $r$ which do not satisfy the above equation the DBI action becomes imaginary, indicating a tachyonic instability \cite{Burgess:1986dw}.

The equations of motion for $g(r)$ and $\theta(r)$ reduce to\footnote{The dilaton in the DBI action has no dependence on $\theta$ and $\phi$ as it is in the solutions we know from string and M-theory.} (We
assume that the components of metric are independent of $\phi(r,t)$.) 
\bse\begin{align}
 \label{A1} A&=\frac{\partial {\cal{L}}}{\partial g'(r)} = \frac{- g_{\phi\phi} (g_{tt} g_{xx} + E^2) g'(r)}{{\cal{L}}}, \\
 \label{EOMthetad2} 0&=\partial_r\left(\frac{g_{\theta
 \theta}\big[(2\pi\alpha')^2E^2 + g_{xx}
 (g_{tt}+\omega^2g_{\phi\phi})]\theta'(r)}{{\cal{L}}}\right)-\frac{\partial{\cal{L}}}{\partial\theta},
\end{align}\ese %
where $A$ is an arbitrary constant. Using \eqref{A1}, $g'(r)$ can be written as
\be\label{eomphi}%
 g'(r)=\sqrt{\frac{c(r)}{d(r)}},
\ee %
where %
\bse\begin{align} %
\label{c1} c(r)&=-(g_{rr}+g_{\theta\theta}\theta'(r)^{2})(g_{tt}g_{xx}+\omega^2g_{xx}g_{\phi\phi}+(2\pi\alpha')^2
E^2)A^2, \\
\label{d1} d(r)&=g_{\phi\phi}(g_{tt}g_{xx}+(2\pi\alpha')^2
E^2)(g_{\phi\phi}(g_{tt}g_{xx}+(2\pi\alpha')^2 E^2)+A^2).
\end{align}\ese %
In general this solution can become imaginary. The reality condition on the solution is fulfilled if we set
$c(r_*)=0$ and $d(r_*)=0$ simultaneously. Finding $r_*$ from
$c(r_*)=0$ and putting it back into $d(r_*)=0$ leads to the desired $A$.
Note that the first parenthesis in \eqref{c1} does not vanish and therefore $r_*$ is given by %
\be\label{rstar} %
 \bigg(g_{tt}g_{xx}+\omega^2g_{xx}g_{\phi\phi}+(2\pi\alpha')^2E^2\bigg)_{r=r_*}=0.
\ee %
According to the reality condition on the action \eqref{reality} the first paranthesis in \eqref{d1} does not become zero and hence from
$d(r_*)=0$ the value of $A$ is given by
\be\label{eqA} %
A=\sqrt{-g_{\phi\phi}(g_{tt}g_{xx}+(2\pi\alpha')^2 E^2)},
\ee %
where all functions of $r$ are evaluated at $r_*$. Interestingly,
the condition \eqref{reality} implies the reality of both the action
and $A$. So far we have defined everything in a general form and
independent of coordinates. In order to obtain the quantities such
as $r_{*}$, $A$ and the temperature one needs to specify the precise
form of the background metric as we will do in the following
sections.\\

~~~
\hspace{-1.2cm}{\tiny{$\blacksquare$}}~ \textbf{Poincare coordinate}\\
~~~
Let's assume the background metric to be $AdS_5\times S^5$ in Poincare' coordinate
\be %
 g_{rr}=\frac{1}{r^2},\ -g_{tt}=g_{xx}=r^2,\ g_{\phi\phi}=\sin^2\theta(r),\
 g_{\theta\theta}=1.
\ee %
The constants $A$ and $r_{*}$ in this coordinate are
\bse\begin{align}\label{rstard2} %
 r_{*}&=\sqrt{\frac{\omega^2\sin^2\psi+\sqrt{4(2\pi\alpha')^2E^{2}+\omega^4\sin^4\psi}}{2}}, \\
 \label{A-D2-P}A&=\sqrt{(r_{*}^{4}-(2\pi\alpha')^2E^{2})\sin^2\psi},
\end{align}\ese %
where $\theta(r_{*})=\psi$. To find the temperature according to \eqref{temperature} we need
to find the horizon by solving \eqref{horizon} which yields to
\be\label{rhrstar} %
r_{h}=\omega \sin\theta(r_h).
\ee %
One can easily see that if the electric field is set to zero $r_*$ and $r_h$ coincide but generally $r_*$ sits outside the horizon.

From \eqref{horizon} and \eqref{temperature} it can be seen that in order to obtain the temperature one needs to know the solution to the equation of motion for $\theta(r)$.  But this equation is a second order differential equation and two initial conditions are needed to solve it. These conditions can be set to be $\psi=\theta(r_*)$ and $\beta=\theta'(r_*)$. In order to find $\beta$ one can expand the equation of motion for $\theta(r)$ around $r_*$ and try to solve it. This can be done numerically but it's not possible to analytically read the value of $\theta(r_h)$ from it. Therefore we are not able to give an explicit form for the temperature for a general from of $\theta(r)$ and will consider the two analytically solvable cases which are $\theta(r)=\frac{\pi}{2}$ with $E\neq0$ and $\theta(r)$ with $E=0$.

\begin{itemize}
\item\textbf{Temperature in the presence of electric field for} {\boldmath{$\theta(r)=\frac{\pi}{2}$}}\\
It is clearly seen from \eqref{EOMthetad2}, $\theta(r)=\frac{\pi}{2}$ is a solution of
the equation of motion. For this solution $r_h=\omega$ and the
temperature is
\be\label{tempthetapi} %
 T=\frac{\omega}{2\pi}\sqrt{\frac{(1-e^2)
(2e^2+\sqrt{1+4e^2}-1)}
 {e^2(\sqrt{1+4e^2}+1)}},
\ee %
where $e=\frac{(2\pi\alpha')E}{\omega^{2}}$. In the zero electric
field limit, the temperature becomes%
\be\label{D2TEB0)}%
T=\frac{\sqrt{2} \omega}{2 \pi},
\ee%
which is
in agreement with the result of \cite{Das:2010yw}. The behaviour of the
temperature with respect to $E$ is shown in figure \ref{Temp of
D2-E}(left) for various values of $\omega$. As it is clear from this figure and equation \eqref{tempthetapi}, there is a maximum value for the electric field. The reason comes from the reality condition on the action
\eqref{reality} which for $AdS_5 \times S^5$ in Poincare' coordinate states that $r$ must be larger than $\sqrt{2 \pi \alpha' E}$. The equation \eqref{tempthetapi} is valid only for the values of electric field that are smaller than $\frac{\omega^2}{2\pi\alpha'}$ since the horizon is at $\omega$.  

\begin{figure}[ht]
\begin{center}
\includegraphics[width=2.6 in]{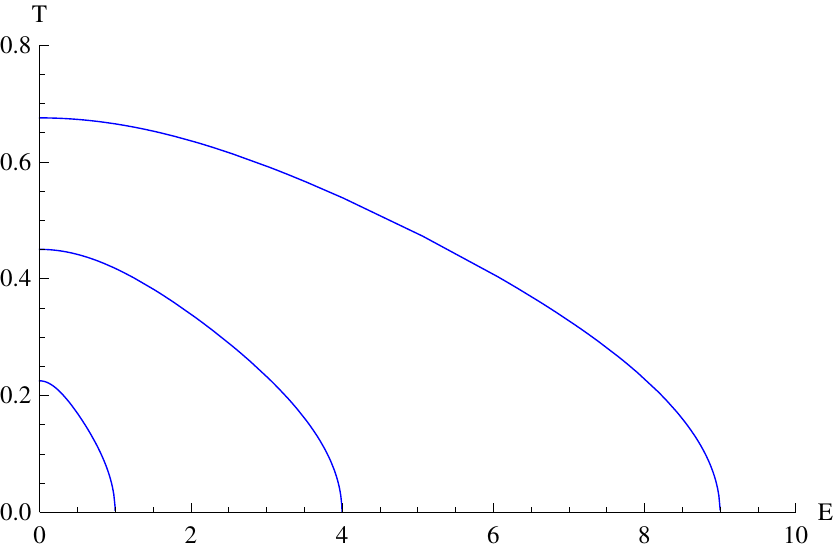}
\hspace{2mm}
\includegraphics[width=2.6 in]{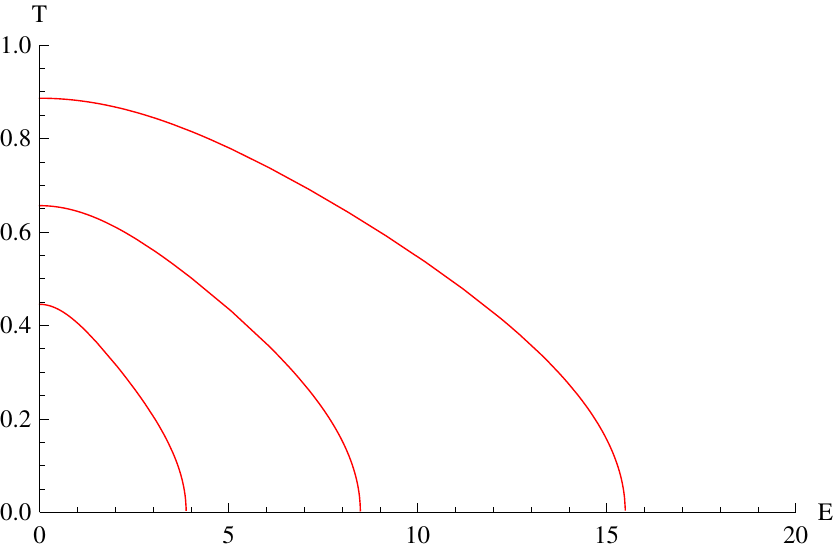}
\caption{The temperature of the induced metric on the D2-Brane in the presence of electric
field is plotted for $\theta(r)=\frac{\pi}{2}$ and $2\pi\alpha'=1$. The
left diagram is plotted in the Poincare' coordinate for $\omega=1, 2, 3$
(bottom to top) and the right diagram is plotted in the global
coordinate for $\omega=2.1, 3, 4$ (bottom to top). \label{Temp of
D2-E}}
\end{center}
\end{figure}%

\item\textbf{Temperature in the absence of electric field}\\
In this case \eqref{rstard2} indicates that  $r_*=r_h=\omega\sin\psi$. Hence the
temperature will be
\be %
T=
\frac{\omega\sin\psi}{2\pi}\sqrt{\frac{(1-\beta\omega\cos\psi)(2+\beta\omega\cos\psi)}{\beta^2\omega^2\sin^2\psi+1}},
\ee %
where $\beta=\theta'(r_*)$. To compute $\beta$, consider the $\theta(r)$ expansion near
$r=r_*$
\be\label{thetaexpansion} %
\theta(r) = \psi + \beta (r-r_*) + \frac{1}{2} \gamma (r-r_*)^2 + ....
\ee %
Substituting the above equation into \eqref{EOMthetad2}, the equation of motion at zeroth order automatically vanishes
and the first order equation yields to an equation for $\beta$ in terms of $\psi$ and $\omega$. Solving this equation we get
\be\label{betad2} %
\beta=\frac{(3-\sqrt{10-6\cos2\psi}-\cos2\psi)\csc^2\psi\sec\psi}{2\omega}.
\ee %
Note that for given $\psi$ between $0$ and $\pi/2$,
$\beta\omega$ is negative and smaller than one. We then have a real
value for the temperature. For $\psi\rightarrow\frac{\pi}{2}$,
$\beta$ is equal to zero and hence the temperature reduces to \eqref{D2TEB0)}. The behaviour of the temperature with respect to $\psi$ is shown in figure \ref{Temp of D2}(left).
\begin{figure}[ht]
\begin{center}
\includegraphics[width=2.6 in]{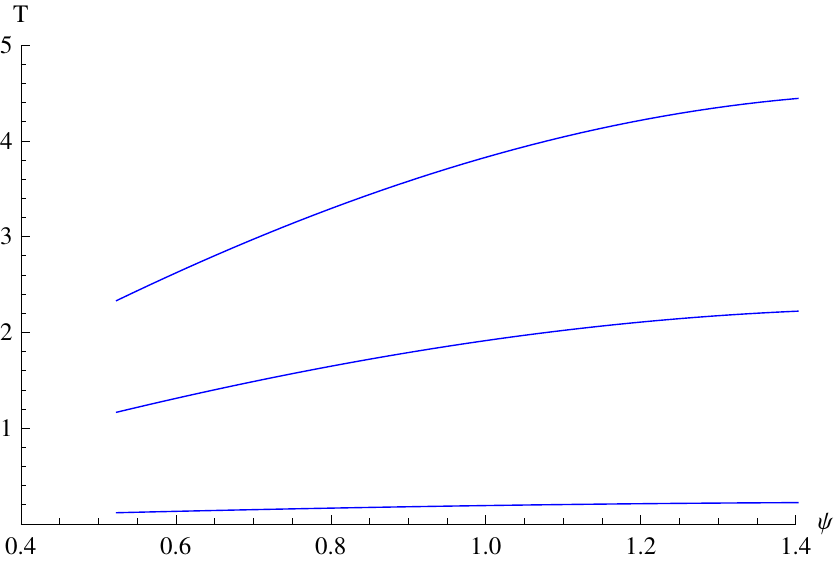}
\hspace{2mm}
\includegraphics[width=2.6 in]{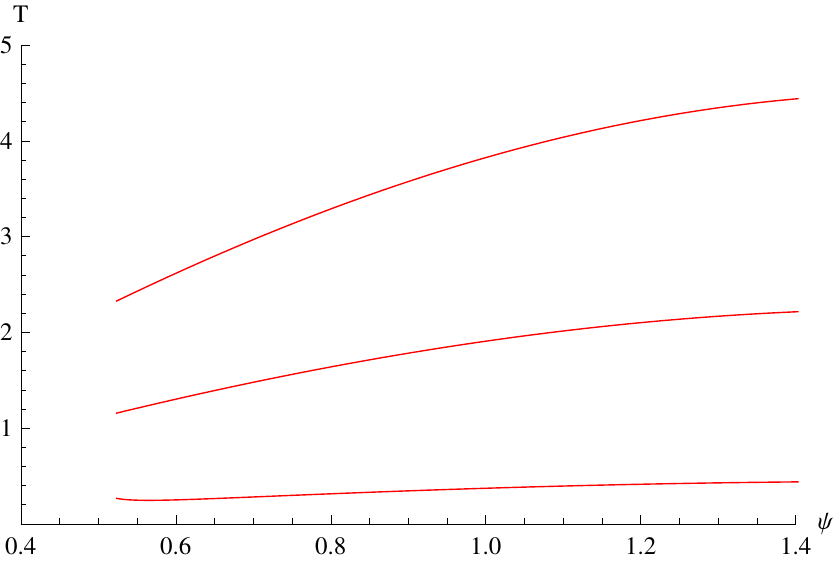}
\caption{The temperature of the induced metric on the D2-Brane in the absence of electric
field is plotted for $\psi$ between $\frac{\pi}{6}$ and
$\frac{\pi}{2}$ and $2\pi\alpha'=1$. Left diagram is in the Poincare'
coordinate for $\omega=1, 2, 3$ (bottom to top) and right diagram is
in the global coordinate for $\omega=2.1, 10, 20$ (bottom to top)
\label{Temp of D2}}
\end{center}
\end{figure}%
\end{itemize}

~~~
\hspace{-1.2cm}{\tiny{$\blacksquare$}}~ \textbf{Global coordinate}\\
~~~
Now let us consider the case of $AdS_5\times S^5$ in global coordinate
\be\label{global coordinate} %
 ds^2=-(1+r^2) dt^2+\frac{dr^2}{1+r^2}+r^2 d\tilde{\Omega}_3^2+d\theta^2+\sin^2\theta d\phi^2+\cos^2\theta d\Omega^2_{3},
\ee%
where %
\bse\begin{align}
 d\tilde{\Omega}_3^2=d\theta_1^2+\sin^2\theta_1 d\phi_1^2+\cos^2\theta_1 d\psi_1^2.
\end{align}\ese %
Note that in this case the D2-brane is extended along $t,r$ and $\theta_1$. We also turn on an electric field as 
\be %
 F_{t1}=E\ dt\wedge e^{(1)},
\ee %
where %
\be %
 e^{(1)}=d\theta_1,\ e^{(2)}=\sin\theta_1d\phi_1,\ \ e^{(3)}=\cos\theta d\psi_1.
\ee %
Using \eqref{rstar} and \eqref{eqA}, $r_*$ and $A$ for a rotating D2-brane in the above background becomes %
\bse\begin{align}\label{rstard2global} %
 r_{*}&=\sqrt{\frac{1}{2}(\omega^2\sin^2\psi-1)+\frac{1}{2}\sqrt{4(2\pi\alpha')^2E^2+(\omega^2\sin^2\psi-1)^2}}, \\
 \label{A-D2-G} A&=\sqrt{r_*^2(1+r_*^2)-(2\pi\alpha')^2E^2}\sin\psi.
\end{align}\ese %
The reality condition \eqref{reality} guaranties that $A$ is real.
In addition due to the electric field, $r_*$ and
$r_h=\sqrt{\omega^2\sin^2\theta(r_h)-1}$ are not coincident and a horizon
exist only when $|\omega\sin\theta(r_h)|>1$. 
Similar to the case of Poincare' coordinate we consider the two
following options that they can be solved explicitly.
\begin{itemize}
\item\textbf{Temperature in the presence of electric field for} {\boldmath{$\theta(r)=\frac{\pi}{2}$}}\\
In this case the temperature is easily read as %
\be
T=\frac{1}{2\pi}\sqrt{\frac{(\omega^2-1)[\frac{1}{\omega^2-1}-e^2][2e^2+\omega^2(\sqrt{4e^2+1}-1)]}{e^2\omega^2(1+\sqrt{4e^2+1})}},
\ee %
where $e=\frac{(2\pi\alpha')E}{\omega^2-1}$. The figure \ref{Temp of
D2-E}(right) shows the value of the temperature as a function of electric field.
As discussed before the upper limit for the electric field comes from the reality condition \eqref{reality}.

\item\textbf{Temperature in the absence of electric field}\\
The temperature on the worldvolume of the probe brane in this case turns out to be
\be %
 T=\frac{(\omega^2\sin^2\psi-1)^{1/4}}{4\pi}\sqrt{\frac{(2-\frac{\beta\omega^2\sin2\psi}{\sqrt{\omega^2\sin^2\psi-1}})(\beta\omega^2\sin2\psi
 +\frac{4\omega^2\sin^2\psi-2}{\sqrt{\omega^2\sin^2\psi-1}})}{\beta^2\omega^2\sin^2\psi+1}}.
\ee %
Similar to the analogy discussed before the temperature can be calculated which in terms of $\psi$ has been plotted in figure \ref{Temp of D2}(right).
\end{itemize}

An interesting quantity one can study in this system is the rate of change of the
energy density. In order to compute it, we need to calculate the
stress-energy tensor of the D$p$-brane which is done in
\cite{Karch}. According to the results of this paper, the
expectation value of the dual field theory stress-energy tensor is
given by
\be %
 \langle T^a_{\ b} \rangle=\int dr \Theta^a_{\ b},
\ee %
where
\be %
 \Theta^a_{\ b}={\cal{L}}\delta_{\ b}^a+2F_{c b}\frac{\delta\cal{L}}{\delta
 F_{ac}}-\partial_b\phi\frac{\delta {\cal{L}}}{\delta\partial_a\phi},
\ee %
is the stress-energy tensor density of the D$p$-brane and $S_{Dp}=\int dr {\cal{L}}$. Therefore the time evolution of the total energy is %
\be %
 \langle \partial_t T_{tt}\rangle=-\langle \partial_t T^t_{\ t}\rangle=-\int_{0}^{\infty} dr\partial_t\Theta^t_{\ t}
 =\int_{0}^{\infty} dr\partial_r\Theta^r_{\ t}=\Theta^r_{\ t}|_{r=0}^{r=\infty}.
\ee %
In the case of D2-brane it is easy to see that $\Theta^r_{\ t}=-\omega\frac{\partial {\cal{L}}}{\partial \phi'(r)}$ and using \eqref{A1} we have $\Theta^r_{\ t}=-\omega A$. Therefore 
\be\label{emt2} %
 \langle \partial_t T_{tt}\rangle=-\omega A|_{r=0}^{r=\infty}=0,
\ee %
where $A$ has been introduced in \eqref{A-D2-P} or \eqref{A-D2-G} for different coordinates.
The above result shows that the total energy is conserved but the
flux of the energy at the boundary and the $r=0$ is non-zero and
equal. In other words an external source injects energy into the
system at the boundary and the same value of energy is dissipated
into the center of the $AdS$ background at $r=0$. On the gauge
theory side, this means that the flavour(thermal) sector injects
energy into the gluon(zero-temperature) sector. As long as the probe
limit is valid we can ignore the dissipation rate. It was discussed
in \cite{Hoyos:2011us} that for times larger than $N_c/\lambda$ the
probe limit is not a reliable approximation and after this time a
black hole may form in the center of $AdS$ background or
equivalently the gluon sector may thermalize.

\subsection{D3-brane}
In this section we will study the effect of external electric and
magnetic fields on the temperature where the probe brane is D3-brane.
It is extended along directions $t,r,x\equiv x_1$ and $y\equiv
x_2$ in the background \eqref{background} . We assume the same ansatz for transverse directions, $\phi(r,t)$ and $\theta(r)$, as before. In contrast to the D2-brane case we can have both constant electric and magnetic fields turned on here. Therefore the action describing the dynamics of
the D3-brane is given by \be\begin{split}
 {\cal{L}}&=\bigg{[}-\left((2\pi\alpha')^2g_{yy}E^2
 +g_{tt}(g_{xx}g_{yy}+(2\pi\alpha')^2B^2)\right)g_{\phi\phi} g'(r)^2\cr
 &-(g_{rr}+g_{\theta
 \theta}\theta'(r)^2)\left((2\pi\alpha')^2g_{yy}E^2 +
 (g_{tt}+\omega^2g_{\phi\phi})\left(g_{xx}
 g_{yy}+(2\pi\alpha')^2B^2\right)\right)\bigg{]}^{\frac{1}{2}},
\end{split}\ee %
where $E=F_{tx}$ and $B=F_{xy}$. The reality condition on the
action becomes %
\be\label{reality2} %
 |g_{tt}|g_{xx}g_{yy}>(2\pi\alpha')^2(g_{yy}E^2-|g_{tt}|B^2).
\ee %
Similar to the calculations done for the D2-brane the equation
of motion for $\phi(r,t)$ leads to the following form for
$g'(r)=\sqrt{\frac{c(r)}{d(r)}}$
\be\label{NU}\begin{split} %
c(r)=&-A^2\left(g_{rr}+g_{\theta\theta}\theta'(r)^2\right)\cr
&\times\big[(2\pi\alpha')^2g_{yy}E^2+(g_{tt}+\omega^2g_{\phi\phi})\left(g_{xx}g_{yy}+(2\pi\alpha')^2B^2
\right)\big], %
\end{split}\ee
\be\begin{split} %
d(r)=&g_{\phi
\phi}\big[(2\pi\alpha')^2g_{rr}E^2+g_{tt}(g_{xx}g_{yy}+(2\pi\alpha')^2B^2)\big]\cr %
&\times\big[A^2+(2\pi\alpha')^2g_{yy}g_{\phi\phi}E^2
+g_{tt}g_{\phi\phi}\left(g_{xx}g_{yy}+(2\pi\alpha')^2B^2\right)\big].
\end{split}\ee %
Therefore from the reality condition on the solution we get
\be\label{A}%
A=\sqrt{-g_{\phi\phi}\big[g_{tt}g_{xx}g_{yy}+(g_{yy}E^2+g_{tt}B^2)\big]}\ |_{r=r_*}.
\ee %
Because of \eqref{reality2} $A$ is real.
Also the equation of motion for $\theta(r)$ is %
\be\label{EOMtheta} %
\partial_r\left(\frac{g_{\theta
 \theta}\big[(2\pi\alpha')^2g_{yy}E^2 +
 (g_{tt}+\omega^2g_{\phi\phi})\left(g_{xx}
 g_{yy}+(2\pi\alpha')^2B^2\right)\big]\theta'(r)}{{\cal{L}}}\right)-\frac{\partial{\cal{L}}}{\partial\theta}=0.
\ee %

In the presence of an external electric field
$r_*$ and $r_h$ do not coincide and the same problem mentioned in
the D2-brane case will rise. Conversely, when only the magnetic field is turned
on we still can have $r_* = r_h$. Therefore we will calculate the
temperature for two different cases. One is where the electric field
is zero but there exists a nonzero constant magnetic field on the
probe brane. And the other one is where we assume
$\theta(r)=\frac{\pi}{2}$ (a solution to the equations of motion)
and electric field is nonzero.
\\

~~~
\hspace{-1.2cm}{\tiny{$\blacksquare$}}~ \textbf{Poincare coordinate}\\
~~~
In the following we consider various
possibilities which can analytically be solved.

\begin{itemize}
 \item \textbf{Temperature in the presence of external magnetic field}\\
 By solving $c(r_*)=0$, $r_*$ turns out to be
 \be %
 r_*=\omega\sin\psi,
 \ee %
 which is the same as $r_h$ obtained from the induced metric. Note that the value of $r_*(=r_h)$ is independent of the magnetic field.

 Using \eqref{temperature}, the temperature of probe brane can easily be computed as
 \be %
 T=\frac{\sqrt{\frac{\omega^2 (\beta\omega\cos\psi-1)\sin\psi
 \left(2\omega^4(3+\beta\omega\cos\psi)\sin^5\psi+(2\pi\alpha')^2B^2(2\sin\psi+\beta\omega\sin(2\psi))\right)}
 {\left(1+\beta^2\omega^2\sin^2\psi\right)\left((2\pi\alpha')^2B^2+\omega^4\sin^4\psi\right)}}}{2 \sqrt{2} \pi}.
 \ee %
 In figure \ref{T-B} the temperature as a function of external
 magnetic field has been plotted.  The main result, obviously seen from this
 figure, is that the temperature reaches a constant value for large
 values of external magnetic field, independent of $\psi$.

\begin{figure}[ht]
\begin{center}
\includegraphics[width=2.6 in]{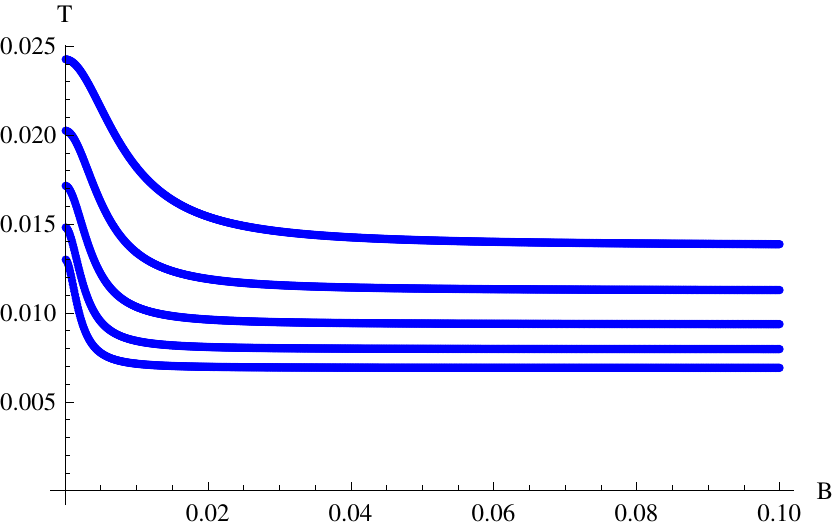}
\hspace{2mm}
\includegraphics[width=2.6 in]{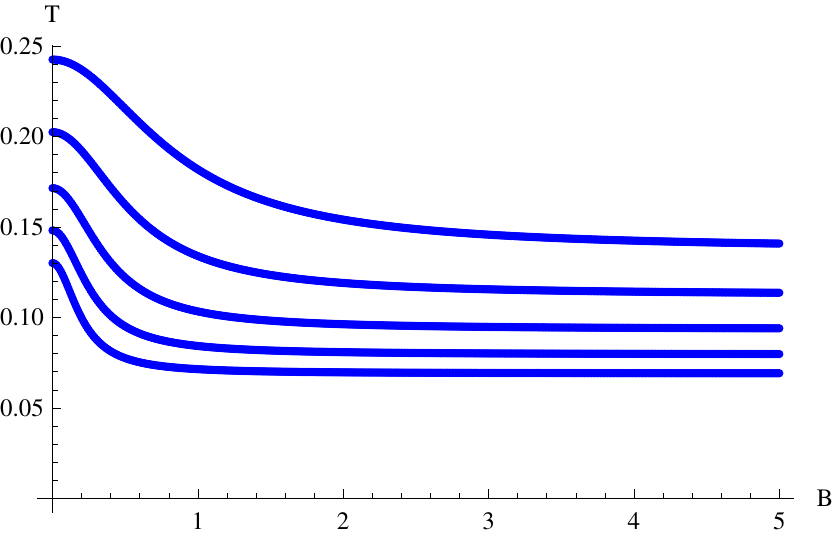}
\caption{The value of temperature as a function of external magnetic
field in Poincare' coordinate has been plotted for $\psi=\frac{\pi}{3}, \frac{\pi}{4},
\frac{\pi}{5}, \frac{\pi}{6}, \frac{\pi}{7}, \frac{\pi}{8}$ (top to
bottom) and $2\pi\alpha'=1$. We set $\omega=0.1$($\omega=1$) in the
left(right) figure.\label{T-B}}
\end{center}
\end{figure}%

A specific solution for the equation of motion \eqref{EOMtheta} is
$\theta(r)=\frac{\pi}{2}$. For this solution $\beta$ is equal to zero
and as a result the temperature becomes
\be %
 T=\frac{\omega}{2\pi}\sqrt{\frac{3
 \omega^{4}+(2\pi\alpha')^{2}B^{2}}{\omega^{4}+(2\pi\alpha')^{2}B^2}}.
\ee %
Therefore in the large magnetic field limit the temperature reaches
\be %
 T=\frac{\omega}{2\pi}.
\ee %
Conversely, in the limit $B\rightarrow 0$ the temperature becomes
\be\label{ccc} %
 T=\frac{\sqrt{3}\omega}{2\pi},
\ee %
which is in agreement with the result of \cite{Das:2010yw}.

\item \textbf{Temperature in the presence of external electric field}\\
Similar to the case of D2-brane, we have to choose $\theta(r)=\frac{\pi}{2}$ and then the temperature in Poincare coordinates is %
\be\label{bbb} %
T=\frac{\omega\sqrt{\omega^4-(2\pi\alpha')^2E^2}}{\sqrt{2}\pi}\frac{\sqrt{\omega^2\sqrt{\omega^4+4(2\pi\alpha')^2E^2}+4(2\pi\alpha')^2E^2-\omega^4}}
{(2\pi\alpha')E(\omega^2+\sqrt{\omega^4+4(2\pi\alpha')^2E^2})}.
\ee %
In the limit in which
$E\rightarrow 0$, \eqref{bbb} becomes \eqref{ccc}.

\item \textbf{Temperature in the absence of electric and magnetic fields}\\
Again similar to the case of D2-brane, in this case the final result for the temperature is
\be %
T=\frac{\omega\sin\psi}{2\pi}\sqrt{\frac{(1-\beta\omega\cos\psi)(3+\beta\omega\cos\psi)}{\beta^2\omega^2\sin^2\psi+1}}.
\ee %

\end{itemize}

\hspace{-1.2cm}{\tiny{$\blacksquare$}}~ \textbf{Global coordinate}\\

Here we consider a rotating D3-brane extended along $t,\ r,\ \theta_1$ and $\phi_1$ in the global coordinate  \eqref{global coordinate}. The ansatz for the electric and magnetic fields in this coordinate is \cite{Filev:2012ch}
\be %
 F_{t1}=E\ dt\wedge e^{(1)},\ \ \ F_{12}=B\ e^{(1)}\wedge e^{(2)}.
\ee %
Now we discuss the following cases:
\begin{itemize}
\item\textbf{Temperature in the presence of external magnetic field}\\
In this coordinate the equations \eqref{horizon} and $c(r_*)=0$ lead to %
\be %
 r_h=r_*=\sqrt{\omega^2\sin^2\psi-1},
\ee %
where \eqref{NU} is used. We can compute the temperature and $\beta$ using \eqref{temperature}
and the equation of motion for $\theta(r)$. However, they are not
illuminating enough so that we will not present them here. In figure \ref{T-Bg}
for different values of $\psi$ the temperature in terms of magnetic
field has been plotted.
\begin{figure}[ht]
\begin{center}
\includegraphics[width=2.6 in]{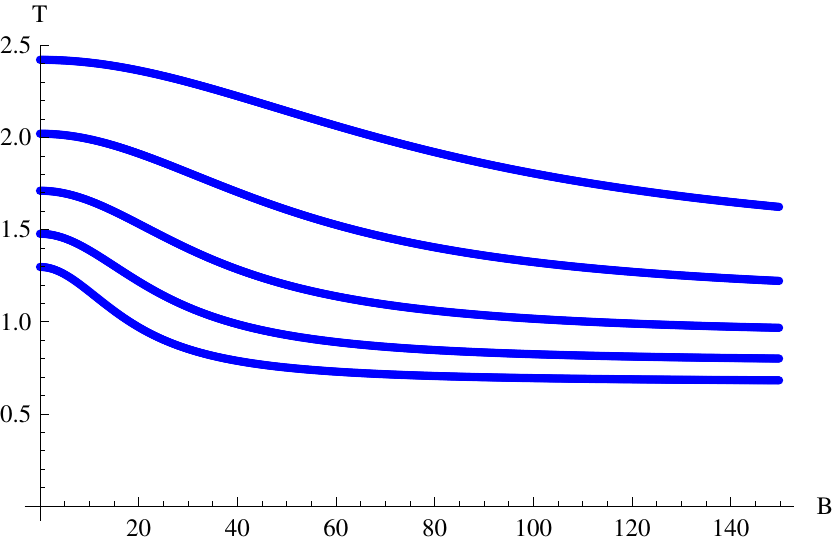}
\hspace{2mm}
\includegraphics[width=2.6 in]{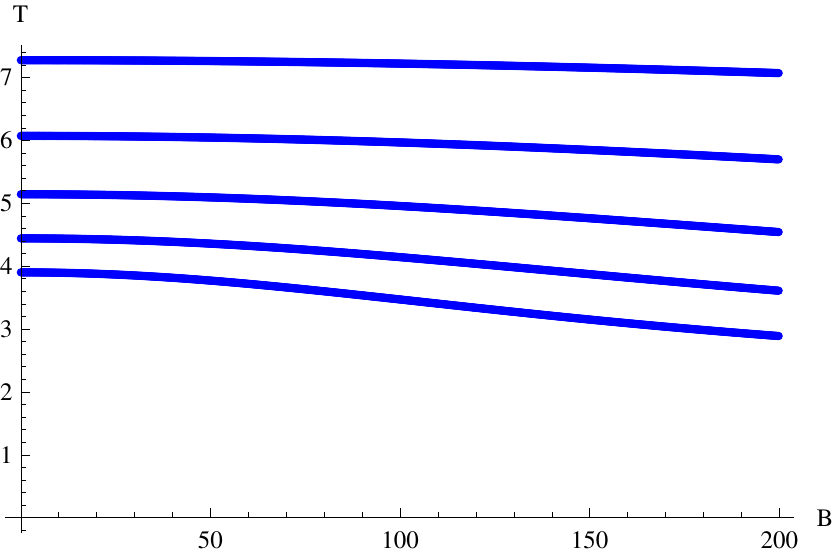}
\caption{The value of temperature as a function of external magnetic field in global coordinate has been plotted
for $\psi=\frac{\pi}{3}, \frac{\pi}{4}, \frac{\pi}{5}, \frac{\pi}{6}, \frac{\pi}{7}, \frac{\pi}{8}$ (top to bottom) and $2\pi\alpha'=1$.
We set $\omega=10$($\omega=30$) in the left(right) figure.\label{T-Bg}}
\end{center}
\end{figure}%

For $\theta(r)=\frac{\pi}{2}$, $\beta$ is zero and hence we have %
\be %
T=\frac{\sqrt{\frac{\left(\omega^2-1\right) \left(1+(2\pi\alpha')^2B^2-4\omega^2+3\omega^4\right)}
{(2\pi\alpha')^2B^2 +\left(\omega^2-1\right)^2}}}{2\pi }.
\ee %
For large values of magnetic field the temperature reaches $\frac{\sqrt{\omega^2-1}}{2\pi}$.
Moreover in the limit of zero magnetic field, the temperature becomes $\frac{\sqrt{3 \omega^2-1}}{2\pi}$ \cite{Das:2010yw}.
\end{itemize}
We can also discuss the cases where the electric field is nonzero for the solution $\theta=\frac{\pi}{2}$ and for a general profile for $\theta(r)$ where we assume $E=B=0$ in global coordinates. The results are quantitatively different but qualitatively the same. Therefore we will not present them here. Moreover the discussion
about the energy-momentum tensor can also be extended to D3-brane. We just need to replace $A$ in \eqref{emt2} by $A$ calculated in the Poincare coordinate or in the global coordinate.

\section{Worldvolume horizon on the rotating D7-brane}
In this section we continue to evaluate the temperature for the probe D7-brane which is embedded in the background \eqref{background}. We choose the bulk to be $AdS_5\times S^5$ in Poincare' coordinate that comes from the near
horizon geometry of a stack of D3-branes. The schematic form of this embedding is
\be %
\begin{array}{cccccccccc}\label{graph}
   & t & r & x_1 & x_2 & x_3 & \Omega_3 & \theta & \phi &  \\
D3 & \times & \times & \times & \times & \times & & &  \\
D7 & \times & \times & \times & \times &  \times  &  \times &  &
\end{array}
\ee %
In contrast to the previous studies of D2 and D3-branes, $\theta=\frac{\pi}{2}$ does not specify the solution to the $\phi(t,r)$ equation of motion for a general ansatz for transverse directions of the form \eqref{configuration}. One can study the shape of the rotating D7-brane and obtain the classical solution for $\theta(r)$ and $\phi(t,r)$. This has been done in \cite{Das:2010yw} and the temperature coming from the induced metric on the probe brane for different profiles has been evaluated. We would like to generalize this calculation to nonzero external electric and magnetic field. 

\begin{figure}[ht]
\begin{center}
\includegraphics[width=2.6 in]{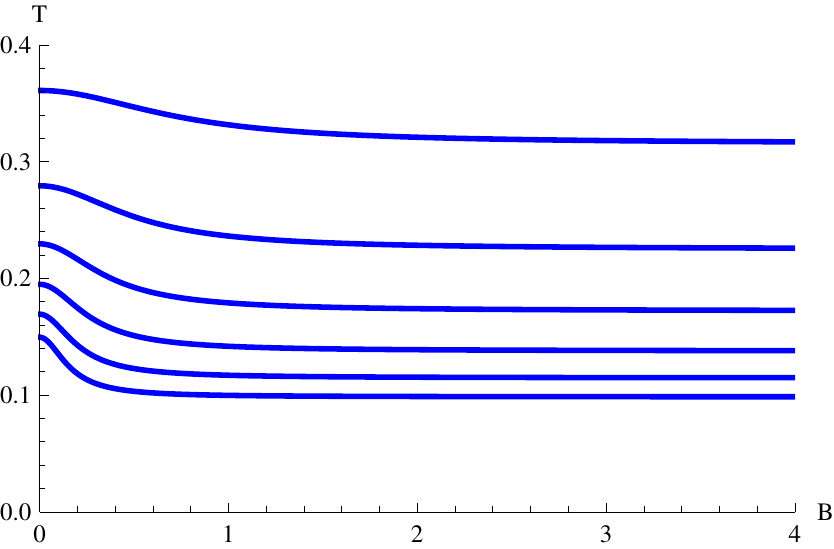}
\hspace{2mm}
\includegraphics[width=2.6 in]{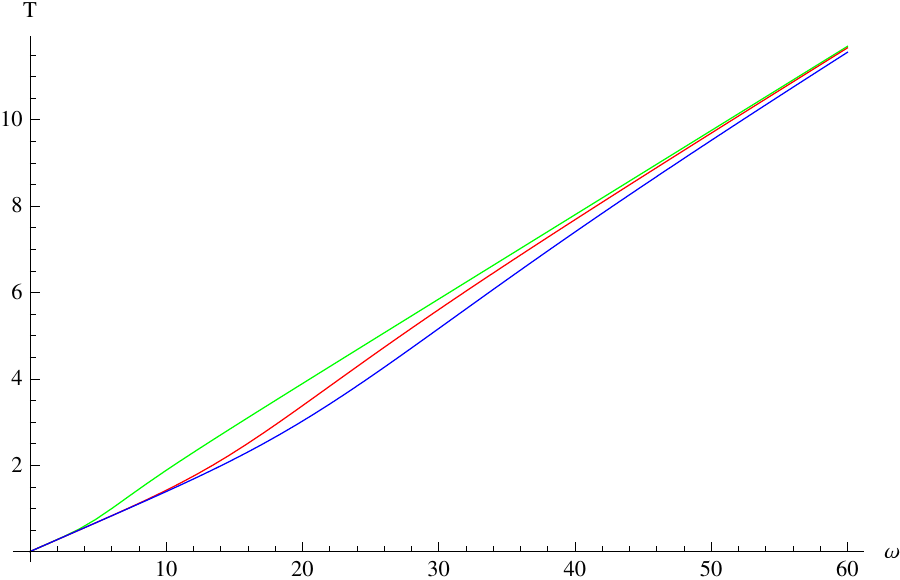}
\caption{Left: The value of temperature as a function of external magnetic field for $\psi=\frac{\pi}{3}, \frac{\pi}{4}, \frac{\pi}{5}, \frac{\pi}{6}, \frac{\pi}{7}, \frac{\pi}{8}$ (top to bottom). We have set $\omega=1$. \newline
Right: The value of temperature as a function of $\omega$ for $B=10$(green), $B=100$(red) and $B=200$(blue). We have chosen $\psi=\frac{\pi}{6}$ and $2\pi\alpha'=1$.\label{T-BD7}}
\end{center}
\end{figure}

To begin we assume the magnetic field to be nonzero on the D7-brane
with the ansatz $F_{x_1 x_2} = B dx_1\wedge dx_2 $. The Lagrangian coming from the
DBI action reduces to %
\bea %
 {\cal{L}} &=& r \cos^3\theta(r) \sqrt{r^4+ (2 \pi \alpha')^2 B^2}\cr
 &\times& \sqrt{\left(r^2 - \omega^2
 \sin^2\theta(r)\right) \left(\frac{1}{r^2} + {\theta'(r)}^2\right) + r^2 \sin^2\theta(r){g'(r)}^2}.
\eea %
Similar to the previous studies the
solution to the equation of motion for $\phi(r,t)$ leads to %
\be
g'(r) =\frac{A \csc \theta(r) \sqrt{(r^2 \theta'(r)^2+1)  (\omega^2\sin^2\theta(r)-r^2)}}{r^2 \sqrt{A^2-r^4 (r^4+(2 \pi
\alpha')^2 B^2) \sin^2\theta(r) \cos^6\theta(r)}}, \ee %
where from
the reality condition on the solution we get \bse\begin{align}
r_* &= \omega \sin\psi, \\
A &= r_*^2 \cos^3\psi \sin\psi \sqrt{r_*^4+ (2 \pi \alpha')^2 B^2}.
\end{align}\ese
Replacing this solution into the induced metric shows that the horizon is at $r=r_*$ and $\theta(r)=\psi$. Therefore one can read the temperature from the induced metric using \eqref{temperature}. The form of the result is not illuminating so we don't write it here.
We have plotted the dependence of the temperature on magnetic field $B$ and angular velocity $\omega$ in figure \ref{T-BD7}. The plot on the left shows how the temperature changes with $B$ for different values of $\psi$. One can see that the behaviour of the temperature is the same as D3-branes where it decreases with increasing the value of the magnetic field until in reaches a constant value (Compare the figure \ref{T-BD7}(left) with \ref{T-B}.). Such behaviour is independent of the value of $\psi$.  The constant value that $T$ reaches at large $B$ is %
\be%
\lim_{B\rightarrow \infty} T=\frac{\omega}{4 \pi }  \sqrt{4-6 \cos 2 \psi +\sqrt{2} \sqrt{(7-8 \cos 2 \psi+3 \cos 4 \psi )} \sec \psi }.
\ee%
It is interesting that this value depends on $\omega$ linearly. Note that for $0 \langle \psi \langle\frac{\pi}{2}$ the function under the square root is always positive. 

In the other plot, figure \ref{T-BD7}(right), we can explicitly see that the temperature increases as the angular velocity is raised. The graphs are plotted for different values of the magnetic field. Interestingly in the limit where $\omega$ is large, the temperature behaves linearly in $\omega$ and is independent of the value of the magnetic field. The value the temperature reaches at large $\omega$ is%
\be%
\lim_{\omega\rightarrow\infty} T = \frac{\omega} {2 \sqrt{2} \pi } \sqrt{3-4 \cos 2 \psi +\sqrt{5-4 \cos 2 \psi } \sec \psi }.
\ee%
Note that, similar to the previous result, for $0 \langle \psi \langle\frac{\pi}{2}$ the function under the square root is always positive.

We can also study the thermalization in the flavour sector in the
presence of the  electric field which is added to the Lagrangian by
$F_{tx_1}=E~dt\wedge dx_1$. Therefore we get %
\bea\label{Ld7E} %
&&{\cal{L}} = r^2 \cos^3\theta(r)\times  \\
&& \sqrt{(r^4 - (2 \pi \alpha')^2 E^2) (\frac{1}{r^2} + {\theta'(r)}^2+\sin^2\theta {g'(r)}^2) -r^2 \omega^2 \sin^2\theta (\frac{1}{r^2} + {\theta'(r)}^2)}.\nonumber \eea %
As before one can impose the reality condition on the Lagrangian. It puts a lower bound on $r$
so that it has to be larger than $\sqrt{2 \pi \alpha' E}$. The classical solution
to the $\phi(t,r)$ equation of motion is %
\bea %
g'(r)=\frac{A \csc \theta  \sqrt{r^2 {\theta'}^2+1}  \sqrt{ r^4 - (2
\pi \alpha')^2 E^2-r^2 \omega^2 \sin^2 \theta}}{r \sqrt{(r^4-(2
\pi \alpha')^2 E^2) (r^4 (r^4-(2 \pi \alpha')^2 E^2) \sin
^2\theta \cos^6\theta-A^2)}},
\eea %
where from the reality condition on the solution we have  
\be %
 A = r_*^2 \sqrt{r_*^4 - (2 \pi \alpha')^2 E^2} \sin\psi \cos^3\psi,
\ee %
where %
\be %
r_* = \sqrt{\frac{\omega ^2 \sin^2\psi + \sqrt{4
 (2 \pi \alpha')^2 E^2+\omega ^4 \sin^4\psi}}{2}}.
\ee %
Note that in this case similar to
the previous cases the horizon is at $r_h = \omega \sin \theta(r_h)$ and
is different from $r_*$. 

In order to obtain the temperature one needs to evaluate the classical solution for $\theta(r)$ at the horizon. In the case of nonzero magnetic field the horizon and $r_*$ coincide. Therefore to find the temperature it was sufficient to calculate $\beta$. In contrast, in the presence of the electric field the horizon is at $\omega \sin \theta(r_h)$ which is different from $r_*$. So one needs to solve the $\theta(r)$ equation of motion for all values of $r$. This can not be done analytically. 
\begin{figure}[ht]
\begin{center}
\includegraphics[width=3.3 in]{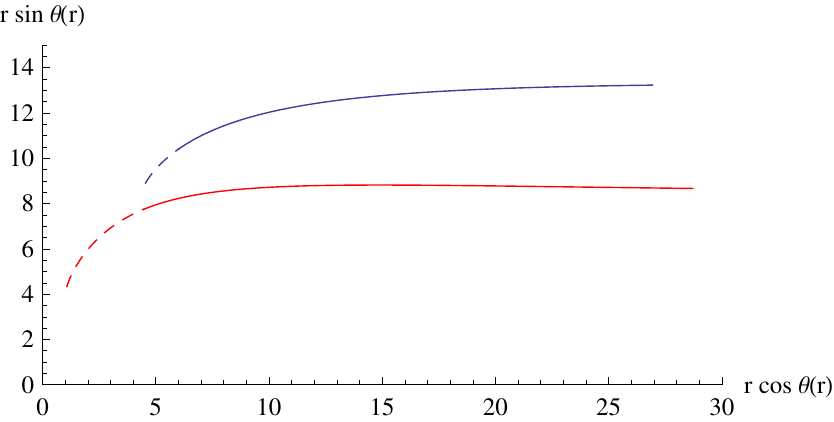}
\caption{The shape of the D7-brane for different values of the electric field where
$\psi=\frac{\pi}{3}, 2\pi\alpha'=1$ and $\omega=10$.\newline
Red: $E=20,\ r_*=8.94427,\ r_{min}=4.47214,\ r_h=8.74453$.\newline
Blue: $E=100,\ r_*=12.0125,\ r_{min}=10$, no horizon.
\label{D7shape}}
\end{center}
\end{figure}%

In order to solve the equation of motion for $\theta(r)$ numerically we need to fix initial conditions which are set by expanding
$\theta(r)$ near $r_*$ as \eqref{thetaexpansion}. 
$\beta$ is functionally dependent on
$E$, $\psi$ and $\omega$. It gets a complicated shape so that we
won't write it here. The equation of motion for $\theta(r)$ is
second order and giving $\psi$ and $\beta$ will be enough to solve
it. Then we obtain numerically
$\theta(r)$ for different values of the external electric field. The numerical solution for $\theta(r)$ has been shown in figure \ref{D7shape} for two values of the electric field. The cross point between the dashed and continuous lines is $r_*$. $r_{min}$ comes from the reality condition of the action as mentioned before. 

The induced metrics corresponding to various solutions to $\theta(r)$ might have a horizon or not. If $\theta(r)$ satisfies the equation $r_h=\omega \sin \theta(r_h)$ the induced metric has a horizon and one can obtain the corresponding temperature. The existence of a horizon depends on how the parameters such as $\omega$, $\psi$ and $E$ are tuned. In figure \ref{D7shape} two configurations with and without horizon have been plotted. For solutions to $\theta(r)$ without a horizon, the flavour sector is not thermalized.  

\begin{figure}[ht]
\begin{center}
\includegraphics[width=2.5 in]{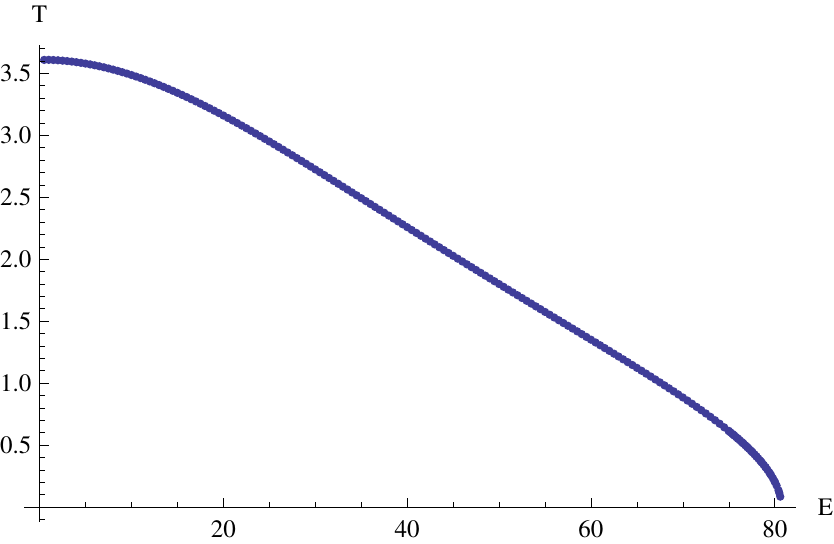}
\hspace{1.5mm}
\includegraphics[width=2.5 in]{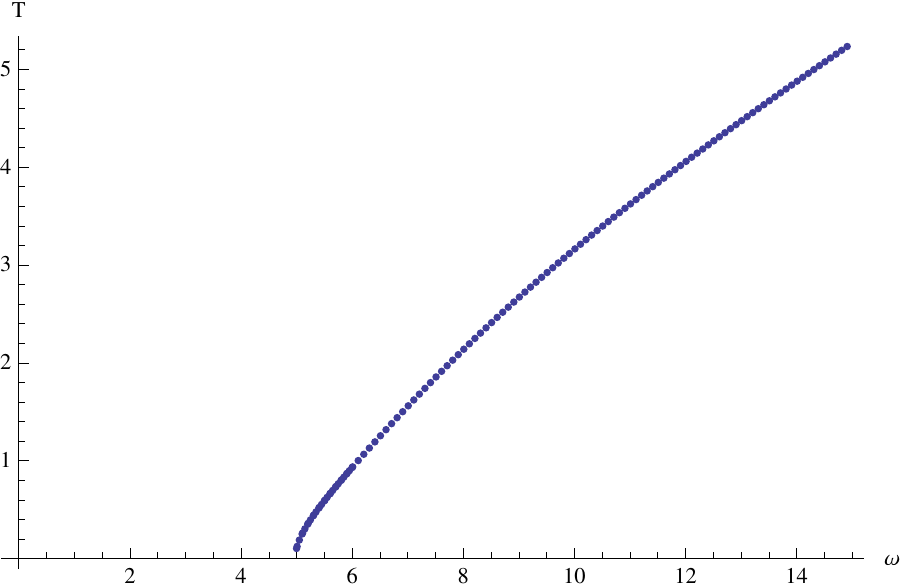}
\caption{Temperature with respect to electric field for $\omega=10$ (left) and with respect to $\omega$ for $E=20$ (right). We choose $2 \pi \alpha'=1$ and $\psi=\frac{\pi}{3}$. \label{D7TE}}
\end{center}
\end{figure}

Following the same analogy discussed in the previous sections we can
read the temperature using \eqref{temperature} for the solutions with horizon. The
dependence of the temperature on the electric field
is shown in figure \ref{D7TE}(left).
It decreases and finally goes to zero as we increase the value of the
external electric field. Interestingly the behaviour of the temperature is similar to the results of D2 and D3-branes where we considered $\theta(r)$ to be constant and equal to $\frac{\pi}{2}$. This can be easily seen by comparing the plot in figure \ref{D7TE}(left) with \ref{Temp of D2-E}. This behaviour of the temperature shows that for fixed values of $\psi$ and  $\omega$ as one increases the electric field the flavour sector temperature decreases until it will not thermalize anymore. 

The behaviour of the temperature with respect to $\omega$ for fixed values of $E$ and $\psi$ has been plotted in figure \ref{D7TE}(right). One can observe that as the angular velocity decreases the temperature decreases until it reaches zero. This means that if there exists a nonzero electric field on the probe brane the rotation in one of the sphere directions can not cause the flavour sector to thermalize until the angular velocity reaches a minimum value specified by the values of $E$ and $\psi$. We have also observed that as $\omega$ increases to much larger values the temperature increases almost linearly in $\omega$. We have checked this behaviour up to $\omega=300$ for the choice of parameters described in figure \ref{D7TE}(right). In fact this behaviour is similar to the behaviour of the temperature in the presence of nonzero magnetic field shown in figure \ref{T-BD7}(right).

\section{Thermal Fluctuations}\label{fluctuationsection}
So far we have been discussing how the electric and magnetic fields affect the thermalization on the probe brane. The probe brane is rotating in one of its transverse directions. The rotation causes the induced metric on the brane to have a horizon although the background is at zero temperature. This means that the flavour sector gets thermalized with the temperature obtained from the induced metric. 

In the presence of nonzero electric field one can define an effective temperature coming from the open string metric which describes the effective geometry seen by open strings \cite{Kim:2011qh,Sonner:2012if}. The open string metric is defined as%
\be%
{\hat{g}}_{ab} = G_{ab} - (F G^{-1} F)_{ab},
\ee%
where $G_{ab}$ is the induced metric on the brane. If we calculate it for the D2-brane set-up explained in section \ref{D2brane}, where we choose the solution $\theta(r)=\frac{\pi}{2}$, after the following change of coordinate
\be%
d\tau=dt- \omega^2\sqrt{\frac{r^2r_*^2}{(r^4-(2\pi\alpha')^2E^2-r^2\omega^2)(r^4-(2\pi\alpha')^2E^2)(r^4-r_*^4)}}\ dr,
\ee%
 we get%
\bea%
d{\hat{s}}^2&=& -\frac{1}{r^2}(r^2-r_*^2)(r^2-r_-^2)d\tau^2 + \frac{r^4-(2\pi\alpha')^2E^2}{r^2(r^4-r_*^4)} dr^2\cr%
\cr
&+& \left(r^2-\frac{(2\pi\alpha')^2 E^2 \left(1+r^2 g'(r)^2\right)}{r^2-\omega ^2+r^4 g'(r)^2}\right) dx^2,
\eea%
where $r_-^2=\frac{1}{2}(\omega^2-\sqrt{\omega^4+4(2\pi\alpha')^2E^2})$ and $g'(r)$ is introduced in \eqref{eomphi}.
One can explicitly see that, in contrast to the horizon of the induced metric, the open string metric horizon lies at $r=r_*$ given by \eqref{rstard2}. Therefore the temperature reads%
\be\label{osmT}%
T_{\text{open}}=\frac{1}{2\pi }\sqrt{\frac{4(2\pi\alpha')^2E^2}{\omega^2}+\omega^2+\sqrt{4(2\pi\alpha')^2E^2+\omega^4}},
\ee%
where in the limit $E\rightarrow 0$ reduces to \eqref{D2TEB0)}. An intriguing observation is that $T_{\text{open}}$ is always larger than \eqref{tempthetapi} which is the temperature obtained from the induced metric before. Intuitively the difference in the temperature can be explained in the following way. The nonzero temperature of the flavour sector coming from the induced metric results in thermal pair production in the flavour sector. On the other hand the nonzero electric field on the probe brane can help to destabilize meson bound states as it pulls the charged particles apart. Therefore the produced pairs result in a current that causes the fundamental matter to feel a hot wind and effectively a different temperature \cite{Kim:2011qh,Liu:2006nn}.

The open string metric in fact appears in the action written for the fluctuations of the probe brane in the directions transverse to it. We
assume the transverse coordinates have the following form%
\be%
x^I(\xi)=x_0^I(\xi)+\delta x^I(\xi),
\ee%
where $\xi$ refers to the coordinates on the probe brane. By expanding the DBI action we get%
\be\begin{split}\label{expansionaction} %
S_{DBI}&= S_0 + S_1 + ... \cr &=- \mu_2\int d^3\xi\sqrt{-\gamma_0} - \frac{\mu_2}{2} \int d^{3}\xi
\sqrt{-\gamma_0}~\bigg{(} \gamma_0^{ab} M_{ba} + \gamma_0^{ab}
N_{ba}\cr &\hspace{3.5 cm}-\frac{1}{2} \gamma_0^{ab}
M_{bc}\gamma_0^{cd} M_{da}\ + \frac{1}{4} (\gamma_0^{ab} M_{ba})^2\bigg{)} +
...,
\end{split}\ee%
where%
\be%
\gamma_{0~ab} = G_{ab} + G_{IJ} \partial_a x_0^I \partial_b x_0^J + (2 \pi \alpha') F_{ab}.
\ee%
$M_{ab}$ and $N_{ab}$ include terms first order and second order in the fluctuations $\delta x^I$, respectively \cite{AliAkbari}. In the case of the D2-brane we would like to study the scalar fluctuations along the transverse direction $\theta$%
\be%
\label{fluctuation}
\theta(t,r,x)=\frac{\pi}{2} + \delta\theta(t,r,x).
\ee%
After replacing the above ansatz into \eqref{expansionaction} the action for fluctuations reduces to%
\begin{equation}
S_1=-\frac{\mu_2}{2}\int d^3\xi \left(\frac{\det G}{\det{\hat{g}}}\right)^\frac{1}{4} \sqrt{-\det{\hat{g}}}\ \hat{g}^{ab}\partial_a\delta\theta\partial_b\delta\theta.
\end{equation} 
It can be seen that the metric observed by these fluctuations is the open string metric up to a scaling factor. If we write the action in the canonical form or the action of a minimally coupled scalar the temperature felt by the fluctuations along the direction $\theta$ is%
\begin{equation}
T_{\text{fluc}}=\frac{1}{2\pi}\sqrt{4\pi^2 T_{open}^2+\frac{(2\pi\alpha')^2E^2\omega^2}{4(2\pi\alpha')^2E^2+\omega^4}},
\end{equation}
which again reduces to \eqref{D2TEB0)}. An interesting observation is that this temperature is different from the one previously obtained from the open string metric \eqref{osmT}. In fact due to the rotation and nonzero electric field the system is out of equilibrium. Therefore it is reasonable to think that different constituents of the flavour sector feel different temperatures. 

\section*{Acknowledgement}

M. A. and H. E.  would like to thank A. O'Bannon and U. A. Wiedemann for fruitful discussions. They also thank CERN theory
division for their hospitality during their visits. M. A. also thanks A. E. Mosaffa for useful discussions.

\end{document}